\documentclass[aps,prl,reprint,amsmath,amssymb,nofootinbib,floatfix]{revtex4-1}
\usepackage{graphicx} 
\usepackage{physics}
\usepackage{letltxmacro}
\usepackage{hyperref}
\usepackage{verbatim}
\usepackage{ulem}
\usepackage{isomath}
\usepackage{cleveref}

\renewcommand{\Re}{\real}
\renewcommand{\Im}{\imaginary}

\usepackage[dvipsnames]{xcolor}

\newcommand{\inst}[1]{\boldsymbol{\mathcal{#1}}}

\LetLtxMacro{\originaleqref}{\eqref}
\renewcommand{\eqref}{Eq.~\originaleqref}

\newcommand{\tens}[1]{\boldsymbol{\mathsf{{#1}}}}

\date{\today}

\begin{document}

\title{Non-linear bistability in pulsed optical traps}

\author{Alex J. Vernon}
\affiliation{Department of Physics and London Centre for Nanotechnology, King's College London, Strand, London WC2R 2LS, UK}

\author{Francisco J. Rodr\'iguez-Fortu\~no}
\affiliation{Department of Physics and London Centre for Nanotechnology, King's College London, Strand, London WC2R 2LS, UK}

\author{Anatoly V. Zayats}
\email{a.zayats@kcl.ac.uk}
\affiliation{Department of Physics and London Centre for Nanotechnology, King's College London, Strand, London WC2R 2LS, UK}

\begin{abstract}
Optical trapping, also known as optical tweezing or optical levitation, is a technique that uses highly focused laser beams to manipulate micro- and nanoscopic particles.
In optical traps driven by high-energy pulses, material non-linearity can result in unusual opto-mechanical effects, such as displaced equilibrium points.
However, existing theoretical models of non-linear optical force on small particles consider smooth material dependence on the incident field strength alone, and not the feedback between the particle permittivity and internal field strength, which is, in turn, a function of the permittivity.
The hysteresis effects of optical bistability in pulsed optical traps therefore elude existing optical force models.
Here, we investigate a bistable optical trap, set up by counter-propagating ultrashort pulses, in which the optical force exerted on a particle depends not only on the field at the particle's current location, but on the particle's historic trajectory in the trap.
The developed formalism will be important for designing optical traps and nanoparticle manipulation in pulsed field for various applications, including potentially time crystal demonstrations.    

\end{abstract}

\maketitle

\section{Introduction}
Focused continuous-wave (CW) lasers are traditionally used to trap and manipulate nanoparticles, molecules and atoms \cite{Ashkin1986,Ashkin1987,Grier2003,Yang2021}.
The prospect of probing unusual nonlinear effects in trapped objects, meanwhile, has drawn much attention to broadband pulse-driven optical traps \cite{Agate2004,Ambardekar2005,Usman2013,Shoji2013,Chiang2014,Goswami2023}.
Under these conditions, no longer is an illuminated particle subject to a continuous optical force, being instead kicked periodically by each passing pulse.
Yet trapping is still possible, assuming that the repetition rate of the pulses is high enough that the particle cannot move a significant distance in the time between two consecutive pulses.
Despite the stroboscopic exposure of trapped particles to what is potentially an extremely large instantaneous optical force, provided non-linear effects are negligible, the average optical force of a pulse is largely similar to its CW counterpart of the same average power \cite{Shane2010}.
Differences generally arise due to linear material properties, such as dispersion, or the particle geometry \cite{duPreezWilkinson2015} which influences reflectivity with respect to the different frequency components that make up the pulse. 

What a typical CW beam cannot achieve, however, is the extreme instantaneous power density in high-energy pulses that can in optical traps elicit a non-linear response.
Some of the strangest light-matter interactions appear in the non-linear regime, where the permittivity of the particle being trapped, treated as a constant in the CW regime, starts to show its dependence on electric field strength \cite{Boyd2020,Zhang2017}.
However, the electric field strength internal to the particle is itself a function of its non-linear permittivity, which creates a feedback loop between the internal field and material parameters.
It is this feedback that enables optical bistability \cite{Boyd2020,Agrawal1979,Xu2023}, where within a certain input power interval there is more than one self-consistent value of relative permittivity $\varepsilon$ that a material can adopt, resulting in a jump in the value of $\varepsilon$ when smoothly increasing or decreasing input power with a hysteresis effect.
While non-linear effects in optical traps, such as split trap sites \cite{Jiang2010,zhang2018}, have previously been reported, existing theoretical models \cite{Jiang2010,zhang2018,Gong2018,Devi2016,Devi2020,Goswami2021,Bandyopadhyay2021,Zhu2024} do not use a self-consistent description of non-linear material permittivity, assuming instead that it has a smooth dependence simply on the incident field strength or intensity without considering the field that physically develops in the particle (the internal field and particle size both being finite in reality).
The manifestation of bistability---and resultant hysteresis effects---in optical force is therefore not well-understood, which we intend to address in this work.
A simple non-linear extension of optical force expressions for point-like particles is insufficient for this purpose, and we instead model a trapped particle with a small but finite size and finite internal field.

For initial context, we begin with a background of optical trapping forces on dipolar particles in time-harmonic fields.
Particles in the Rayleigh regime (size$\;<<\lambda$) scatter incident light in a dipolar manner and experience an optical force that in monochromatic waves can be expressed as a sum of several intuitive terms \cite{NietoVesperinas2010,Golat2023,Toftul2024}, quadratic with respect to the fields and usually dominated by two in particular.
One of these two, called gradient force, normally draws the particles towards regions of higher energy density and into equilibrium near local maxima, enabling optical trapping in tightly focussed beams \cite{Grier2003}.
The other of the two terms is called radiation pressure and inhibits the trapping ability of any single beam, because it pushes particles in the local direction of propagation.
Gradient force sometimes overwhelms radiation pressure even in a single focussed Gaussian beam \cite{Ashkin1986}, though it is possible to altogether eliminate radiation pressure and isolate the effect of gradient force among multiple trapping sites (antinodes) by using counter-propagating beams.
Doing so accentuates one of gradient force's remarkable properties: that its direction depends on the sign of the real part of the particle polarisability $\alpha(\omega)$.
Often, by de-tuning $\omega$ of the trapping beam(s) from the particle resonant frequency, the direction of gradient force can be completely reversed.
Then, then the particle is no longer attracted to intensity maxima, instead being drawn down into regions of low intensity \cite{Nelson2007,Xu2010,Vernon2024}.

There is however another, far more exotic way to manipulate light-matter interactions and re-shape an optical trap, lying in the non-linear regime \cite{Jiang2010}.
It is our purpose in this paper to theorise how non-linear effects could be leveraged to alter the behaviour of optical traps with a self-consistent model of material permittivity.
By approximating non-linearity in two-dimensional simulations, we show that gradient force reversal can be achieved via the spatial distribution of extreme intensity throughout the envelope of an optical trap formed of high-energy pulses, which is the general explanation for split trap sites in pulsed optical traps \cite{Jiang2010,Gong2018}.
Yet optical bistability produces an extraordinary effect.
We simulate a bistable optical trap, where the average force exerted on a particle depends not only on the particle current location in the spatial profile of a trap field, but on its \textit{historic} location too.

The paper is structured as follows.
The next two sections explain the calculation of an optical force map of a focussed pulse, and the approximation of self-consistent non-linear material behaviour.
Numerical simulations of a nonlinear particle in a pulsed optical field are then presented, demonstrating bistable trapping. 

\section{Optical force from ultrashort pulses}
With the presence of many frequency components in an ultrashort pulse, it is not possible to exactly express pulsed optical forces on Rayleigh particles in the same, time-averaged form as in CW beams, not least because the optical force exerted on a dipolar particle is quadratic with respect to the electric and magnetic field phasors.
Though it may be reasonable to treat a pulse as monochromatic if its spectral bandwidth is narrow enough \cite{Gong2018}, we will not make any approximation in this regard.
A time-dependent optical force can in general be calculated using the time-dependent Maxwell stress tensor and kinetic momentum density.
Although a pulse optical force on a particle varies instantaneously, changing in direction and magnitude as the field amplitude changes, its net mechanical interaction is described simply by the final value of momentum imparted to the particle once the pulse has completely passed by.

Consider a single pulse incident on a volume $V$ containing nanoparticles.
As time evolves and the pulse begins to pass through the volume, the total momentum $\mathbf{P}$ that has so far been accumulated collectively by any matter within the domain $V$ is described by the time-integral of the electromagnetic force \cite{Griffiths,duPreezWilkinson2015}
\begin{equation}\label{accumulated_momentum}
\begin{split}
    \mathbf{P}(t)&=\int_0^t\mathbf{F}(t)dt\\&=\int_0^t\left[\oint_{S}\tens{T}(t)\cdot d\mathbf{a}\right]dt-\int_{V}\frac{1}{c^2}\inst{E}(t)\times\inst{H}(t)dV \, .
\end{split}  
\end{equation}
Here, the first term corresponds to the optical momentum that has so far crossed the volume boundary $S$, i.e., the time-integrated flux of the stress tensor $\tens{T}$, given by \cite{Griffiths}
\begin{equation}
    T_{ij}=\varepsilon_0\left(\mathcal{E}_i\mathcal{E}_j-\frac{1}{2}\delta_{ij}|\inst{E}|^2\right)+\mu_0\left(\mathcal{H}_i\mathcal{H}_j-\frac{1}{2}\delta_{ij}|\inst{H}|^2\right) \, ,
\end{equation}
where $\delta_{ij}=1$ for $i=j$ and $\delta_{ij}=0$ otherwise, and where $\inst{E}(t)$ and $\inst{H}(t)$ are the total instantaneous electric and magnetic fields (combining both incident and scattered fields).
The second term in \eqref{accumulated_momentum} accounts for kinetic momentum carried by light within volume $V$ at the instant time $t$.
When eventually the incident pulse and all scattered light leave the domain, the first term of \eqref{accumulated_momentum}---the only to remain non-zero---corresponds to the net momentum transferred during the complete pulse-matter interaction.
An average force $\mathbf{F}_\text{av}$ exerted by a train of pulses can be found by multiplying this overall, single-pulse momentum by the desired pulse repetition rate.
Through \eqref{accumulated_momentum}, the average force $\mathbf{F}_\text{av}$ we calculate is exact and not limited to dipole-only gradient and radiation pressure force interactions.

If there is only one particle present at some position $\mathbf{r}_\text{p}$ in the domain $V$, therefore acquiring all of the momentum lost by the pulse in \eqref{accumulated_momentum}, the average optical force $\mathbf{F}_\text{av}$ it experiences must be in the direction of $\mathbf{P}$.
But when exerted by a strongly focussed pulse which is confined both in time and space (as is necessary for an optical trap), $\mathbf{F}_\text{av}$ depends not only on the particle material properties, but on its location in the spatial profile of the pulse.
A functional optical trap will contain at least one equilibrium point---a position into which streamlines of position-dependent $\mathbf{F}_\text{av}$ terminate and where a particle will experience no overall force.
For a particle of a given size and material, we can plot a force map of the position-dependent average applied optical force, and any equilibrium points (trap sites), by varying the particle location $\mathbf{r}_\text{p}$ and calculating \eqref{accumulated_momentum}.
This is the procedure which we adopt in the simulations, presented later on.

\section{Non-linear response}
\begin{figure*}[t!]
    \centering
    \includegraphics{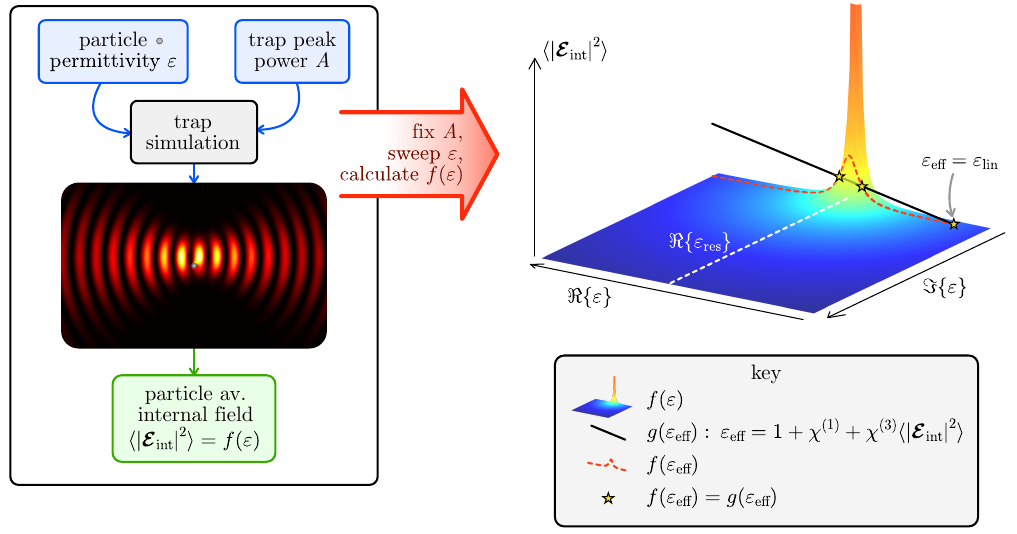}
    \caption{Procedure for determining self-consistent values of the non-linear permittivity of a particle of fixed size at a certain location, in an optical trap formed of fixed energy counter-propagating pulses.
    As the counter-propagating pulses pass over the particle, the duration-averaged internal field that develops $\expval{|\inst{E}_\text{int}|^2}$ is a function $f(\varepsilon)$ of the particle permittivity $\varepsilon$ and is maximal at the particle resonance condition $\varepsilon_\text{res}$.
    If we then choose susceptibilities $\chi^{(1)}$ and $\chi^{(3)}$, describing linear and non-linear optical properties of the particle, then its non-linear permittivity is given by \eqref{non_lin_eps}, which can be rearranged into the form $\expval{|\inst{E}_\text{int}|^2}=g(\varepsilon_\text{eff})$.
    Values of $\varepsilon_\text{eff}$ satisfying $f(\varepsilon_\text{eff})=g(\varepsilon_\text{eff})$ are self-consistent.
    When three solutions to ?? are present, as shown in the right hand plot by the stars, the value of permittivity that is physically realised depends on the history of the particle.
    }
    \label{figure_1}
\end{figure*}
The simplest description of the non-linear response of a material and its dependence on electric field strength is typically provided by an expansion of a scalar polarisation density in the material \cite{Boyd2020},
\begin{equation}\label{P_instantaneous}
    \mathcal{P}(t)=\varepsilon_0\left[\chi^{(1)}\mathcal{E}(t)+\chi^{(2)}\mathcal{E}^2(t)+\chi^{(3)}\mathcal{E}^3(t)+...\right].
\end{equation}
\noindent In centrosymmetric materials, the third-order non-linear susceptibility $\chi^{(3)}$ is the leading non-linear parameter and is often sufficient to define an effective non-linear permittivity $\varepsilon_\text{nl}$ of the material (or a corresponding non-linear refractive index)
via time-averaging of the field strength, denoted by angled brackets:
\begin{equation}\label{non_lin_eps_instantaneous}
    \varepsilon_\text{nl}\approx1+\chi^{(1)}+\chi^{(3)}\expval{\mathcal{E}^2(t)},
\end{equation}
where the sum of the first two terms constitute the material linear permittivity.
Equations (\ref{P_instantaneous})-(\ref{non_lin_eps_instantaneous}) make some significant simplifications: the material is assumed to be lossless and dispersion free, and immersed only in a linearly polarised electric field which is not so strong as to warrant contributions to $\varepsilon_\text{nl}$ from terms with higher-order susceptibilities.


The non-linear material behaviour of a small nanoparticle of finite size and finite internal field can be modelled by adapting \eqref{non_lin_eps_instantaneous}.
We define an effective non-linear permittivity that depends on the magnitude of the internal (non-scalar) electric field in the nanoparticle averaged over the pulse duration:
\begin{equation}\label{non_lin_eps}
    \varepsilon_\text{eff}=1+\chi^{(1)}+\chi^{(3)}\expval{|\inst{E}_\text{int}|^2}.
\end{equation}
Taking the square of $\inst{E}_\text{int}$, we are neglecting the loss of isotropy of the material effective permittivity due to the variable strength of different field components \cite{Boyd2020,Zhu2024}, but \eqref{non_lin_eps} will be sufficient for our purposes in modeling bistability.
Despite the simple appearance of \eqref{non_lin_eps}, to actually determine the effective permittivity that a particle takes on at some position in the spatial envelope of a pulse with a certain energy and duration is rather complicated.
This is because $\expval{|\inst{E}_\text{int}|^2}$ is the duration-averaged field strength \textit{within} the particle, which itself depends on, and therefore must be self-consistent with, the effective particle permittivity $\varepsilon_\text{eff}$.
Many theoretical models \cite{Devi2016,Devi2020,Goswami2021,Bandyopadhyay2021,Zhu2024} of optical force that incorporate field-strength-dependence of permittivity (and therefore particle polarisability) do not stretch beyond an equivalent expression of \eqref{non_lin_eps_instantaneous}, and in doing so appear to assume that the incident field alone determines non-linear permittivity.
To model non-linearity in this way removes the feedback that couples the total field and the particle material parameters, and with it, any notion of optical bistability.
In reaching self-consistency between the variables in \eqref{non_lin_eps}, we adopt a parameter sweep approach described visually in Fig.~\ref{figure_1}.

We consider now a particle of arbitrary permittivity $\varepsilon$ situated somewhere within the spatial envelope of a pulse trap of fixed peak power.
The pulse-duration-averaged electric field strength that physically develops in the particle is itself a function of both real and imaginary parts of $\varepsilon$,
\begin{equation}
    \expval{|\inst{E}_\text{int}|^2}=f(\varepsilon),
\end{equation}
and therefore describes a surface in the $(\Re\{\varepsilon\},\Im\{\varepsilon\},\expval{|\inst{E}_\text{int}|^2})$ parameter space, shown on the right in Fig.~\ref{figure_1}.
Notably, if the particle is small then $f$ has a distinct peak located where $\varepsilon$ meets the particle resonance condition and the internal field diverges. For a small spherical particle in free space, this permittivity is $\varepsilon=-2$, and for a small 2D cylindrical particle under transverse-electric (TE) illumination this is near to $\varepsilon=-1$.
By specifying a non-linear material for the particle with appropriate values of $\chi^{(1)}$ and $\chi^{(3)}$, its permittivity must now satisfy \eqref{non_lin_eps}, which because $\expval{|\inst{E}_\text{int}|^2}$ is a real scalar, spans only a one-dimensional space of possible values of $\varepsilon_\text{eff}$.
That must mean that \eqref{non_lin_eps}, which can be rearranged into the form $\expval{|\inst{E}_\text{int}|^2}=g(\varepsilon_\text{eff})$, is the equation of a straight line in the three-dimensional parameter space of $(\Re\{\varepsilon\},\Im\{\varepsilon\},\expval{|\inst{E}_\text{int}|^2})$, whose direction in the complex $\varepsilon$ plane depends on the value of $\chi^{(3)}$ (see black line in Fig.~\ref{figure_1}).
Any set of parameters satisfying $g(\varepsilon_\text{eff})=f(\varepsilon)$ (the intersection of the straight line $g$ with the surface $f$) corresponds to a self-consistent value of effective permittivity, such as is shown by the stars in Fig.~\ref{figure_1}.

\begin{figure}[t]
    \centering
    \includegraphics[width=\columnwidth]{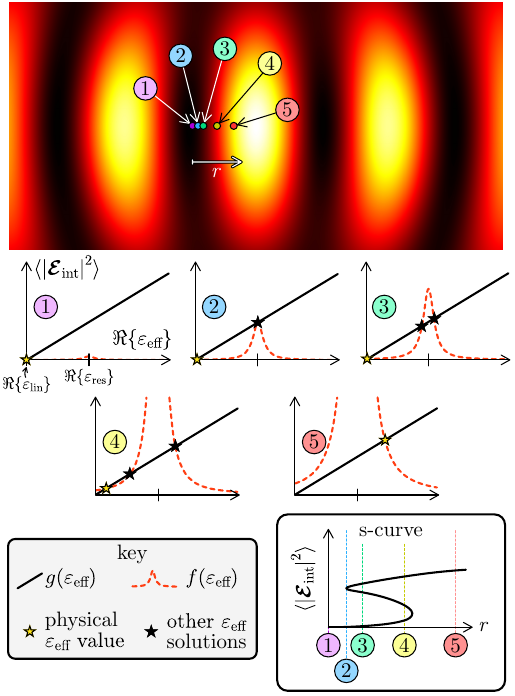}
    \caption{Dependence of the number of self-consistent permittivity values [solutions to $g(\varepsilon_\text{eff})=f(\varepsilon_\text{eff})$] on position of a particle in an optical trap formed of counter-propagating pulses.
    For visual reference of the position of nodes and antinodes in the resultant (time-dependent) standing wave, the energy density in the trap at time $t=0$ is also plotted.
    If a particle smoothly diffuses from a node towards an antinode (position 1 to 5), the electric field strength it is subjected to increases.
    This causes its permittivity to transition from (approximately) the linear permittivity to a permittivity beyond that which corresponds the particle resonance ($\varepsilon_\text{res}$).
    The jump in the physical permittivity of the particle (yellow star) between positions 4 and 5 is characteristic of optical bistability.}
    \label{figure_2}
\end{figure}

It is because of the resonance peak in $f$ that multiple intersections between it and $g(\varepsilon_\text{eff})$ can arise, characteristic of optical bistability.
There are three key factors for a dipolar particle that determine the number of solutions to $g(\varepsilon_\text{eff})=f(\varepsilon_\text{eff})$: the particle's linear permittivity (determining the starting point of the straight line below the surface $f$), the third-order non-linear susceptibility (determining the direction of the line, towards or away from the resonance peak of $f$), and the peak power of the pulse illumination (which deforms the surface $f$).
The spatial confinement of the trap adds another layer of complexity, since the value of $\expval{|\inst{E}_\text{int}|^2}$ that would develop in a particle---and therefore the self-consistent value(s) of $\varepsilon_\text{eff}$ it inherits in the non-linear regime---depend on the particle position.
Only near the antinodes of the counter-propagating pulse trap is the field strong enough that $\varepsilon_\text{eff}$ can deviate substantially from the linear permittivity and acquire multiple self-consistent values.


With straightforward topological arguments, it can be shown that, as long as $\varepsilon_\text{eff}$ depends linearly on $\expval{|\inst{E}_\text{int}|^2}$, only one or three self-consistent values of $\varepsilon_\text{eff}$ can stably occur for a small particle.
In Fig.~\ref{figure_2}, the curves described by $g(\varepsilon_\text{eff})$ and $f(\varepsilon_\text{eff})$ are plotted for various positions of a particle between two antinodes in a counter-propagating pulse trap, which without changing the input power of the trap subjects the particle to different electric field amplitudes.
For simplicity a positive, purely real value of $\chi^{(3)}$ is chosen (so only $\Re\{\varepsilon_\text{eff}\}$ depends on the electric field) and the real part of the linear permittivity of the particle is negative.
If the particle moves smoothly from a node towards an antinode of the trap, the curve $f(\varepsilon_\text{eff})$ deforms and the number of intersection points of the black and red dashed lines transitions from one, to two (unstably), to three, to one again, via creation and annihilation of solutions to $g(\varepsilon_\text{eff})=f(\varepsilon_\text{eff})$.
This mechanism of creation and annihilation of self-consistent permittivity values explains what would be observed experimentally as bitability.
For, in a real system, a particle can only have one physical permittivity value which initially in Fig.~\ref{figure_2}, when the particle is in a node of the trap, is that nearest to its linear permittivity (indicated by the yellow star).
Only once the particle is in position 5, and only one self-consistent value of $\varepsilon_\text{eff}$ is available, does its permittivity jump to the other side of the resonance peak of $f$.
Had the particle diffused in the opposite direction, from position 5 to position 1, its physical permittivity would undergo a different transition, one which is not the same as the reverse of the transition from position 1 to 5.
The characteristic S-curve tracing the internal field strength versus particle position $r$ (Fig.~\ref{figure_2}) summarises this hysteresis behaviour.
A switching of the particle permittivity to the opposite side of the resonance (determined by the geometry of the small particle) is a possible sign of gradient force reversal which would mean a particle is either attracted to or repelled from the antinodes of the trap, depending on the current and historic positions of the particle in the trap.

\section{Bistable optical traps}
\begin{figure*}[t!]
    \centering
    \includegraphics{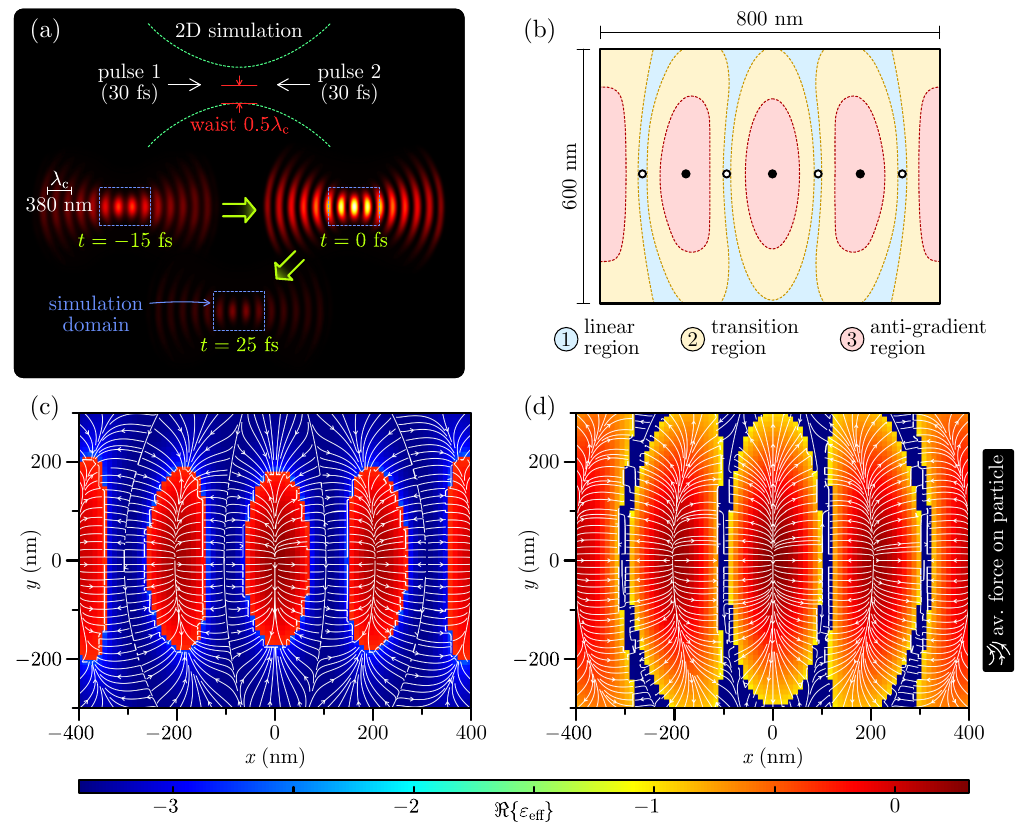}
    \caption{
    Numerical results for a 2D simulation of a material-agnostic cylindrical particle in an optical trap, formed by two counter-propagating pulses, each with an energy-per-length of $0.0012$ J/m.
    The simulation set-up is shown in (a).
    The cylindrical particle diameter is $20$ nm, has a linear permittivity of $\varepsilon_\text{lin}=-3.4+0.2i$, and has a real third-order non-linear susceptibility of $\chi^{(3)}=2\times10^{20}\text{ m}^2/\text{V}^2$.
    Due to the spatial distribution of field intensity in the resulting standing wave, the trap is divided into three regions indicated in (b).
    The impulse accrued by the particle from the pulses, which reverses in the field antinodes, depends both on the current position of the pulse in the trap, and its historic position.
    This dependence is incorporated into two different force maps (c) and (d), where in (c) the particle historic location is in the region of the linear optical behaviour, while in (d) its historic location is in the anti-gradient region.
    \label{figure_3}
    }
\end{figure*}

We simulated a two-dimensional, pulsed optical trap formed by counter-propagating, tightly focussed pulses, having a Gaussian shape both in space and time, that are linearly polarised in the $xy$ plane of the simulation [Fig.~\ref{figure_3}(a)].
A two-dimensional simulation environment was chosen so that the optical trap spatial behaviour can be interpreted easily while reducing computational demands.
We therefore modeled a two-dimensional particle: an infinite cylinder, whose infinite length stretches along the $z$ axis---the $z$ axis invariance of the entire system means that all integrated quantities (e.g., force/impulse on the particle, pulse peak power, pulse energy) are calculated per unit length.
The internal and external scattered fields of the cylinder are calculated using analytic formulae for every plane wave in the angular and frequency spectra of the two pulses forming the trap, according to Ref. \cite{Bohren1998}.
In this way the simulated total fields associated with the particle, having a finite size in the $xy$ plane, are exact up to a chosen order of cylindrical vector harmonics.
This distinguishes our approach from other theoretical models of non-linear optical force, which assume a point-like particle with infinite internal field, whose scattering ability is characterised simply by a scalar or tensor polarisability.

Counter-propagating pulses of centre wavelength $\lambda_\text{c}=380$~nm, 30~fs duration (amplitude FWHM), which are synchronised so as to arrive at their (coincident) focal points at the same time, develop a time-dependent standing wave with the contrast between its nodes and antinodes reaching a maximum at time $t=0$.
A force map of the optical trap was calculated by scanning a cylindrical particle (20~nm diameter) through different positions within the envelope of the trap [bounded by the blue dashed box in Fig.~\ref{figure_3}(a)], and at each position calculating the impulse \eqref{accumulated_momentum} given to the particle as the pulses pass by.
The direction and magnitude of this impulse depends on the material properties of the particle, to which we assigned the effective non-linear permittivity $\varepsilon_\text{eff}$.
It should now be stressed that the purpose of this study is to understand the effects of non-linear material behaviour---particularly bistability---in optical traps and in this respect our results are material-agnostic (implications we shall discuss in the conclusions section).
We modeled the linear permittivity of the particle to be $\varepsilon_\text{lin}=-3.4+0.2i$ and selected a positive, purely real value of $\chi^{(3)}=2\times10^{-20}\text{ m}^2/\text{V}^2$ for the particle to achieve a state of bistability, as described in the previous section.
The order of magnitude of this value of $\chi^{(3)}$ determines the necessary, fixed energy of the pulses for which non-linear behaviour emerges, which in this simulation was $0.0012$ J/m per pulse (recall that this is a 2D simulation, hence we deal with per-unit-length quantities).

With the chosen simulation parameters, we observe position-dependent bistability of the particle permittivity $\varepsilon_\text{eff}$, defining as a consequence three distinct regions in the envelope of the trap [Fig.~\ref{figure_3}(b)]: a linear region, a transition region and an anti-gradient region.
The linear region is a low-intensity region of space, encompassing the nodes of the standing wave, where the electric field is not strong enough to change the particle permittivity from its linear value $\varepsilon_\text{lin}$.
In contrast, the antinodes of the standing wave are contained in the red anti-gradient region [Fig.~\ref{figure_3}(b)], where, remarkably, the electric field is so strong that the particle effective non-linear permittivity jumps beyond that of the particle resonance condition, resulting in a reversal of the direction of gradient force (and therefore the impulse given to a particle situated in that region).
Between the linear and anti-gradient regions is the intermediate transition region, where the field intensity is such that there are three self-consistent values of $\varepsilon_\text{eff}$ of the particle [as explained in Fig.~\ref{figure_2}].
What is special about this transition region is that the impulse imparted by the counter-propagating pulses to a particle changes direction depending on the particle's historic location---specifically, whether the particle had entered the transition region from the linear region or from the anti-gradient region.

Summarising this position-dependent hysteresis are the two numerically simulated force maps of the trap in Fig~\ref{figure_3}(c) and (d).
The background colour of the two plots corresponds to the real part of the particle effective non-linear permittivity, which changes in space as it is exposed to different electric field intensities [note that because we chose a purely real value of $\chi^{(3)}$, it is only the real part of $\varepsilon_\text{eff}$ that changes according to \eqref{non_lin_eps}].
In both (c) and (d), somewhere between the trap nodes and antinodes, there is a sudden change in the particle permittivity (blue to red/orange) as is characteristic of the familiar hysteresis S-curve.
However, the perimeter of this sharp change in permittivity is different between the two plots: in (c), the variation of the particle permittivity is what would occur if the particle diffuses from a node to an antinode, while in (d), the position dependence of $\varepsilon_\text{eff}$ is corresponds to that of a particle diffusing initially from an antinode towards a node.
White streamlines correspond to the position-dependent impulse (proportional to average force) given to the cylindrical particle by the passage of the counter-propagating pulses, and show its direction reversal for a particle located in an antinode compared to a node.

The presence of hysteresis in the opto-mechanics of a non-linear particle raises questions about what we would expect its trajectory to be within the trap, and how it compares to experimental observations.
We anticipate that a large transition region in Fig.~\ref{figure_3}(b) would cause a particle to oscillate in position, as (if gradient force reversal occurs) the optical force direction within the region switches.
But if the transition region in Fig.~\ref{figure_3}(b), which is modulated by the input power and focussing of the pulses, is narrow enough then we could reasonably expect the particle to settle in equilibrium somewhere between the trap nodes and antinodes.

\section{Conclusions}
We have explored an extraordinary possible consequence of non-linear bistability when present in the material behaviour of a particle situated in a counter-propagating, pulsed optical trap.
In a counter-propagating pulse trap of sufficient per-pulse energy, contrasting nodes and anti-nodes are established which divide the spatial envelope of the trap into regions of low electric field strength, where interacting material behaves linearly, and regions of strong intensity where non-linear effects take hold and a trapped particle permittivity is altered substantially from its linear value.
In a typical, linear standing wave trap, a particle would be drawn by gradient force towards equilibrium in the trap antinodes.
But with the right material parameters, it is possible for gradient force to reverse direction due to a change in the particle effective non-linear permittivity brought about by the high intensity at the trap antinodes.
Gradient force reversal due to non-linearity is what has been proposed to explain split trap sites \cite{Jiang2010,zhang2018} in high-power pulse traps, and results in equilibrium points that are displaced from the trap focal point.
However, we have shown the underlying hysteresis of non-linear bistability results in an optical force that, notably, not only depends on the current position of a particle in the trap, but also on its historic location.

The presented 2D simulations were conducted with a material-agnostic cylindrical particle.
There are several factors determining non-linear changes of the permittivity of the particle, including its nonlinear susceptibility, the illuminating pulse duration, the material topology and dispersion of its optical properties.
The scope of this study was to explore how non-linear response of material changes the interaction of a particle and an optical trap.
It should also be noted, therefore, that our chosen material parameters were specific to the cylindrical geometry of the particle in the 2D simulation.
A small spherical particle in 3D real space has a different resonance condition (met at $\varepsilon=-2$), meaning a bistable trap including gradient force reversal has different material and intensity requirements.
This all said, bistability of optical systems is an experimentally verified phenomenon and, regardless of the parameters of a particular bistable non-linear material (e.g., gold), the spatially dependent hysteresis behaviour which is predicted here, where optical force depends on a particle current and historic location within the trap, should still be manifest in some form. 

Going beyond traditional Gaussian pulsed beams to, for example, cylindrical vector beams with complex intensity profiles, may allow achieving bistable behavior and trapping dependent on historic trajectories of multiple non-linear particles, and ultimately emergence of time-crystalline properties in photonic systems.  

\section*{Acknowledgments}
This work was supported by the UK EPSRC project EP/Y015673/1 and the ERC iCOMM project (789340). All the data supporting the findings of this work are presented in the results section and available from the corresponding author upon reasonable request.

\bibliography{bibliography}

\begin{thebibliography}{32}%
\makeatletter
\providecommand \@ifxundefined [1]{%
 \@ifx{#1\undefined}
}%
\providecommand \@ifnum [1]{%
 \ifnum #1\expandafter \@firstoftwo
 \else \expandafter \@secondoftwo
 \fi
}%
\providecommand \@ifx [1]{%
 \ifx #1\expandafter \@firstoftwo
 \else \expandafter \@secondoftwo
 \fi
}%
\providecommand \natexlab [1]{#1}%
\providecommand \enquote  [1]{``#1''}%
\providecommand \bibnamefont  [1]{#1}%
\providecommand \bibfnamefont [1]{#1}%
\providecommand \citenamefont [1]{#1}%
\providecommand \href@noop [0]{\@secondoftwo}%
\providecommand \href [0]{\begingroup \@sanitize@url \@href}%
\providecommand \@href[1]{\@@startlink{#1}\@@href}%
\providecommand \@@href[1]{\endgroup#1\@@endlink}%
\providecommand \@sanitize@url [0]{\catcode `\\12\catcode `\$12\catcode `\&12\catcode `\#12\catcode `\^12\catcode `\_12\catcode `\%12\relax}%
\providecommand \@@startlink[1]{}%
\providecommand \@@endlink[0]{}%
\providecommand \url  [0]{\begingroup\@sanitize@url \@url }%
\providecommand \@url [1]{\endgroup\@href {#1}{\urlprefix }}%
\providecommand \urlprefix  [0]{URL }%
\providecommand \Eprint [0]{\href }%
\providecommand \doibase [0]{http://dx.doi.org/}%
\providecommand \selectlanguage [0]{\@gobble}%
\providecommand \bibinfo  [0]{\@secondoftwo}%
\providecommand \bibfield  [0]{\@secondoftwo}%
\providecommand \translation [1]{[#1]}%
\providecommand \BibitemOpen [0]{}%
\providecommand \bibitemStop [0]{}%
\providecommand \bibitemNoStop [0]{.\EOS\space}%
\providecommand \EOS [0]{\spacefactor3000\relax}%
\providecommand \BibitemShut  [1]{\csname bibitem#1\endcsname}%
\let\auto@bib@innerbib\@empty
\bibitem [{\citenamefont {Ashkin}\ \emph {et~al.}(1986)\citenamefont {Ashkin}, \citenamefont {Dziedzic}, \citenamefont {Bjorkholm},\ and\ \citenamefont {Chu}}]{Ashkin1986}%
  \BibitemOpen
  \bibfield  {author} {\bibinfo {author} {\bibfnamefont {A.}~\bibnamefont {Ashkin}}, \bibinfo {author} {\bibfnamefont {J.~M.}\ \bibnamefont {Dziedzic}}, \bibinfo {author} {\bibfnamefont {J.~E.}\ \bibnamefont {Bjorkholm}}, \ and\ \bibinfo {author} {\bibfnamefont {S.}~\bibnamefont {Chu}},\ }\href {\doibase 10.1364/OL.11.000288} {\bibfield  {journal} {\bibinfo  {journal} {Optics Letters}\ }\textbf {\bibinfo {volume} {11}},\ \bibinfo {pages} {288} (\bibinfo {year} {1986})}\BibitemShut {NoStop}%
\bibitem [{\citenamefont {Ashkin}\ \emph {et~al.}(1987)\citenamefont {Ashkin}, \citenamefont {Dziedzic},\ and\ \citenamefont {Yamane}}]{Ashkin1987}%
  \BibitemOpen
  \bibfield  {author} {\bibinfo {author} {\bibfnamefont {A.}~\bibnamefont {Ashkin}}, \bibinfo {author} {\bibfnamefont {J.~M.}\ \bibnamefont {Dziedzic}}, \ and\ \bibinfo {author} {\bibfnamefont {T.}~\bibnamefont {Yamane}},\ }\href {\doibase 10.1038/330769a0} {\bibfield  {journal} {\bibinfo  {journal} {Nature}\ }\textbf {\bibinfo {volume} {330}},\ \bibinfo {pages} {769} (\bibinfo {year} {1987})}\BibitemShut {NoStop}%
\bibitem [{\citenamefont {Grier}(2003)}]{Grier2003}%
  \BibitemOpen
  \bibfield  {author} {\bibinfo {author} {\bibfnamefont {D.~G.}\ \bibnamefont {Grier}},\ }\href {\doibase 10.1038/nature01935} {\bibfield  {journal} {\bibinfo  {journal} {Nature}\ }\textbf {\bibinfo {volume} {424}},\ \bibinfo {pages} {810} (\bibinfo {year} {2003})}\BibitemShut {NoStop}%
\bibitem [{\citenamefont {Yang}\ \emph {et~al.}(2021)\citenamefont {Yang}, \citenamefont {Ren}, \citenamefont {Chen}, \citenamefont {Arita},\ and\ \citenamefont {Rosales-Guzmán}}]{Yang2021}%
  \BibitemOpen
  \bibfield  {author} {\bibinfo {author} {\bibfnamefont {Y.}~\bibnamefont {Yang}}, \bibinfo {author} {\bibfnamefont {Y.-X.}\ \bibnamefont {Ren}}, \bibinfo {author} {\bibfnamefont {M.}~\bibnamefont {Chen}}, \bibinfo {author} {\bibfnamefont {Y.}~\bibnamefont {Arita}}, \ and\ \bibinfo {author} {\bibfnamefont {C.}~\bibnamefont {Rosales-Guzmán}},\ }\href {\doibase 10.1117/1.AP.3.3.034001} {\bibfield  {journal} {\bibinfo  {journal} {Advanced Photonics}\ }\textbf {\bibinfo {volume} {3}} (\bibinfo {year} {2021}),\ 10.1117/1.AP.3.3.034001}\BibitemShut {NoStop}%
\bibitem [{\citenamefont {Agate}\ \emph {et~al.}(2004)\citenamefont {Agate}, \citenamefont {Brown}, \citenamefont {Sibbett},\ and\ \citenamefont {Dholakia}}]{Agate2004}%
  \BibitemOpen
  \bibfield  {author} {\bibinfo {author} {\bibfnamefont {B.}~\bibnamefont {Agate}}, \bibinfo {author} {\bibfnamefont {C.~T.~A.}\ \bibnamefont {Brown}}, \bibinfo {author} {\bibfnamefont {W.}~\bibnamefont {Sibbett}}, \ and\ \bibinfo {author} {\bibfnamefont {K.}~\bibnamefont {Dholakia}},\ }\href {\doibase 10.1364/OPEX.12.003011} {\bibfield  {journal} {\bibinfo  {journal} {Optics Express}\ }\textbf {\bibinfo {volume} {12}},\ \bibinfo {pages} {3011} (\bibinfo {year} {2004})}\BibitemShut {NoStop}%
\bibitem [{\citenamefont {Ambardekar}\ and\ \citenamefont {qing Li}(2005)}]{Ambardekar2005}%
  \BibitemOpen
  \bibfield  {author} {\bibinfo {author} {\bibfnamefont {A.~A.}\ \bibnamefont {Ambardekar}}\ and\ \bibinfo {author} {\bibfnamefont {Y.}~\bibnamefont {qing Li}},\ }\href {\doibase 10.1364/OL.30.001797} {\bibfield  {journal} {\bibinfo  {journal} {Optics Letters}\ }\textbf {\bibinfo {volume} {30}},\ \bibinfo {pages} {1797} (\bibinfo {year} {2005})}\BibitemShut {NoStop}%
\bibitem [{\citenamefont {Usman}\ \emph {et~al.}(2013)\citenamefont {Usman}, \citenamefont {Chiang},\ and\ \citenamefont {Masuhara}}]{Usman2013}%
  \BibitemOpen
  \bibfield  {author} {\bibinfo {author} {\bibfnamefont {A.}~\bibnamefont {Usman}}, \bibinfo {author} {\bibfnamefont {W.-Y.}\ \bibnamefont {Chiang}}, \ and\ \bibinfo {author} {\bibfnamefont {H.}~\bibnamefont {Masuhara}},\ }\href {\doibase 10.3184/003685013X13592844053451} {\bibfield  {journal} {\bibinfo  {journal} {Science Progress}\ }\textbf {\bibinfo {volume} {96}},\ \bibinfo {pages} {1} (\bibinfo {year} {2013})}\BibitemShut {NoStop}%
\bibitem [{\citenamefont {Shoji}\ \emph {et~al.}(2013)\citenamefont {Shoji}, \citenamefont {Saitoh}, \citenamefont {Kitamura}, \citenamefont {Nagasawa}, \citenamefont {Murakoshi}, \citenamefont {Yamauchi}, \citenamefont {Ito}, \citenamefont {Miyasaka}, \citenamefont {Ishihara},\ and\ \citenamefont {Tsuboi}}]{Shoji2013}%
  \BibitemOpen
  \bibfield  {author} {\bibinfo {author} {\bibfnamefont {T.}~\bibnamefont {Shoji}}, \bibinfo {author} {\bibfnamefont {J.}~\bibnamefont {Saitoh}}, \bibinfo {author} {\bibfnamefont {N.}~\bibnamefont {Kitamura}}, \bibinfo {author} {\bibfnamefont {F.}~\bibnamefont {Nagasawa}}, \bibinfo {author} {\bibfnamefont {K.}~\bibnamefont {Murakoshi}}, \bibinfo {author} {\bibfnamefont {H.}~\bibnamefont {Yamauchi}}, \bibinfo {author} {\bibfnamefont {S.}~\bibnamefont {Ito}}, \bibinfo {author} {\bibfnamefont {H.}~\bibnamefont {Miyasaka}}, \bibinfo {author} {\bibfnamefont {H.}~\bibnamefont {Ishihara}}, \ and\ \bibinfo {author} {\bibfnamefont {Y.}~\bibnamefont {Tsuboi}},\ }\href {\doibase 10.1021/ja401657j} {\bibfield  {journal} {\bibinfo  {journal} {Journal of the American Chemical Society}\ }\textbf {\bibinfo {volume} {135}},\ \bibinfo {pages} {6643} (\bibinfo {year} {2013})}\BibitemShut {NoStop}%
\bibitem [{\citenamefont {Chiang}\ \emph {et~al.}(2014)\citenamefont {Chiang}, \citenamefont {Okuhata}, \citenamefont {Usman}, \citenamefont {Tamai},\ and\ \citenamefont {Masuhara}}]{Chiang2014}%
  \BibitemOpen
  \bibfield  {author} {\bibinfo {author} {\bibfnamefont {W.-Y.}\ \bibnamefont {Chiang}}, \bibinfo {author} {\bibfnamefont {T.}~\bibnamefont {Okuhata}}, \bibinfo {author} {\bibfnamefont {A.}~\bibnamefont {Usman}}, \bibinfo {author} {\bibfnamefont {N.}~\bibnamefont {Tamai}}, \ and\ \bibinfo {author} {\bibfnamefont {H.}~\bibnamefont {Masuhara}},\ }\href {\doibase 10.1021/jp502524f} {\bibfield  {journal} {\bibinfo  {journal} {The Journal of Physical Chemistry B}\ }\textbf {\bibinfo {volume} {118}},\ \bibinfo {pages} {14010} (\bibinfo {year} {2014})}\BibitemShut {NoStop}%
\bibitem [{\citenamefont {Goswami}(2023)}]{Goswami2023}%
  \BibitemOpen
  \bibfield  {author} {\bibinfo {author} {\bibfnamefont {D.}~\bibnamefont {Goswami}},\ }\href {\doibase 10.3389/fchem.2022.1006637} {\bibfield  {journal} {\bibinfo  {journal} {Frontiers in Chemistry}\ }\textbf {\bibinfo {volume} {10}} (\bibinfo {year} {2023}),\ 10.3389/fchem.2022.1006637}\BibitemShut {NoStop}%
\bibitem [{\citenamefont {Shane}\ \emph {et~al.}(2010)\citenamefont {Shane}, \citenamefont {Mazilu}, \citenamefont {Lee},\ and\ \citenamefont {Dholakia}}]{Shane2010}%
  \BibitemOpen
  \bibfield  {author} {\bibinfo {author} {\bibfnamefont {J.~C.}\ \bibnamefont {Shane}}, \bibinfo {author} {\bibfnamefont {M.}~\bibnamefont {Mazilu}}, \bibinfo {author} {\bibfnamefont {W.~M.}\ \bibnamefont {Lee}}, \ and\ \bibinfo {author} {\bibfnamefont {K.}~\bibnamefont {Dholakia}},\ }\href {\doibase 10.1364/OE.18.007554} {\bibfield  {journal} {\bibinfo  {journal} {Optics Express}\ }\textbf {\bibinfo {volume} {18}},\ \bibinfo {pages} {7554} (\bibinfo {year} {2010})}\BibitemShut {NoStop}%
\bibitem [{\citenamefont {du~Preez-Wilkinson}\ \emph {et~al.}(2015)\citenamefont {du~Preez-Wilkinson}, \citenamefont {Stilgoe}, \citenamefont {Alzaidi}, \citenamefont {Rubinsztein-Dunlop},\ and\ \citenamefont {Nieminen}}]{duPreezWilkinson2015}%
  \BibitemOpen
  \bibfield  {author} {\bibinfo {author} {\bibfnamefont {N.}~\bibnamefont {du~Preez-Wilkinson}}, \bibinfo {author} {\bibfnamefont {A.~B.}\ \bibnamefont {Stilgoe}}, \bibinfo {author} {\bibfnamefont {T.}~\bibnamefont {Alzaidi}}, \bibinfo {author} {\bibfnamefont {H.}~\bibnamefont {Rubinsztein-Dunlop}}, \ and\ \bibinfo {author} {\bibfnamefont {T.~A.}\ \bibnamefont {Nieminen}},\ }\href {\doibase 10.1364/OE.23.007190} {\bibfield  {journal} {\bibinfo  {journal} {Optics Express}\ }\textbf {\bibinfo {volume} {23}},\ \bibinfo {pages} {7190} (\bibinfo {year} {2015})}\BibitemShut {NoStop}%
\bibitem [{\citenamefont {Boyd}(2020)}]{Boyd2020}%
  \BibitemOpen
  \bibfield  {author} {\bibinfo {author} {\bibfnamefont {R.~W.}\ \bibnamefont {Boyd}},\ }\href@noop {} {\emph {\bibinfo {title} {Nonlinear Optics}}},\ \bibinfo {edition} {4th}\ ed.\ (\bibinfo  {publisher} {Elsevier},\ \bibinfo {year} {2020})\BibitemShut {NoStop}%
\bibitem [{\citenamefont {xi~Zhang}\ and\ \citenamefont {hua Wang}(2017)}]{Zhang2017}%
  \BibitemOpen
  \bibfield  {author} {\bibinfo {author} {\bibfnamefont {Y.}~\bibnamefont {xi~Zhang}}\ and\ \bibinfo {author} {\bibfnamefont {Y.}~\bibnamefont {hua Wang}},\ }\href {\doibase 10.1039/C7RA07551K} {\bibfield  {journal} {\bibinfo  {journal} {RSC Advances}\ }\textbf {\bibinfo {volume} {7}},\ \bibinfo {pages} {45129} (\bibinfo {year} {2017})}\BibitemShut {NoStop}%
\bibitem [{\citenamefont {Agrawal}\ and\ \citenamefont {Carmichael}(1979)}]{Agrawal1979}%
  \BibitemOpen
  \bibfield  {author} {\bibinfo {author} {\bibfnamefont {G.~P.}\ \bibnamefont {Agrawal}}\ and\ \bibinfo {author} {\bibfnamefont {H.~J.}\ \bibnamefont {Carmichael}},\ }\href {\doibase 10.1103/PhysRevA.19.2074} {\bibfield  {journal} {\bibinfo  {journal} {Physical Review A}\ }\textbf {\bibinfo {volume} {19}},\ \bibinfo {pages} {2074} (\bibinfo {year} {1979})}\BibitemShut {NoStop}%
\bibitem [{\citenamefont {Xu}\ \emph {et~al.}(2023)\citenamefont {Xu}, \citenamefont {Peng}, \citenamefont {Jiang}, \citenamefont {Qian},\ and\ \citenamefont {Jiang}}]{Xu2023}%
  \BibitemOpen
  \bibfield  {author} {\bibinfo {author} {\bibfnamefont {J.}~\bibnamefont {Xu}}, \bibinfo {author} {\bibfnamefont {Y.}~\bibnamefont {Peng}}, \bibinfo {author} {\bibfnamefont {J.}~\bibnamefont {Jiang}}, \bibinfo {author} {\bibfnamefont {S.}~\bibnamefont {Qian}}, \ and\ \bibinfo {author} {\bibfnamefont {L.}~\bibnamefont {Jiang}},\ }\href {\doibase 10.1364/OL.488889} {\bibfield  {journal} {\bibinfo  {journal} {Optics Letters}\ }\textbf {\bibinfo {volume} {48}},\ \bibinfo {pages} {3235} (\bibinfo {year} {2023})}\BibitemShut {NoStop}%
\bibitem [{\citenamefont {Jiang}\ \emph {et~al.}(2010)\citenamefont {Jiang}, \citenamefont {Narushima},\ and\ \citenamefont {Okamoto}}]{Jiang2010}%
  \BibitemOpen
  \bibfield  {author} {\bibinfo {author} {\bibfnamefont {Y.}~\bibnamefont {Jiang}}, \bibinfo {author} {\bibfnamefont {T.}~\bibnamefont {Narushima}}, \ and\ \bibinfo {author} {\bibfnamefont {H.}~\bibnamefont {Okamoto}},\ }\href {\doibase 10.1038/nphys1776} {\bibfield  {journal} {\bibinfo  {journal} {Nature Physics}\ }\textbf {\bibinfo {volume} {6}},\ \bibinfo {pages} {1005} (\bibinfo {year} {2010})}\BibitemShut {NoStop}%
\bibitem [{\citenamefont {Zhang}\ \emph {et~al.}(2018)\citenamefont {Zhang}, \citenamefont {Shen}, \citenamefont {Min}, \citenamefont {Jin}, \citenamefont {Jiang}, \citenamefont {Liu}, \citenamefont {Zhu}, \citenamefont {Sheng}, \citenamefont {Zayats},\ and\ \citenamefont {Yuan}}]{zhang2018}%
  \BibitemOpen
  \bibfield  {author} {\bibinfo {author} {\bibfnamefont {Y.}~\bibnamefont {Zhang}}, \bibinfo {author} {\bibfnamefont {J.}~\bibnamefont {Shen}}, \bibinfo {author} {\bibfnamefont {C.}~\bibnamefont {Min}}, \bibinfo {author} {\bibfnamefont {Y.}~\bibnamefont {Jin}}, \bibinfo {author} {\bibfnamefont {Y.}~\bibnamefont {Jiang}}, \bibinfo {author} {\bibfnamefont {J.}~\bibnamefont {Liu}}, \bibinfo {author} {\bibfnamefont {S.}~\bibnamefont {Zhu}}, \bibinfo {author} {\bibfnamefont {Y.}~\bibnamefont {Sheng}}, \bibinfo {author} {\bibfnamefont {A.~V.}\ \bibnamefont {Zayats}}, \ and\ \bibinfo {author} {\bibfnamefont {X.}~\bibnamefont {Yuan}},\ }\href@noop {} {\bibfield  {journal} {\bibinfo  {journal} {Nano Letters}\ }\textbf {\bibinfo {volume} {18}},\ \bibinfo {pages} {5538} (\bibinfo {year} {2018})}\BibitemShut {NoStop}%
\bibitem [{\citenamefont {Gong}\ \emph {et~al.}(2018)\citenamefont {Gong}, \citenamefont {Gu}, \citenamefont {Rui}, \citenamefont {Cui}, \citenamefont {Zhu},\ and\ \citenamefont {Zhan}}]{Gong2018}%
  \BibitemOpen
  \bibfield  {author} {\bibinfo {author} {\bibfnamefont {L.}~\bibnamefont {Gong}}, \bibinfo {author} {\bibfnamefont {B.}~\bibnamefont {Gu}}, \bibinfo {author} {\bibfnamefont {G.}~\bibnamefont {Rui}}, \bibinfo {author} {\bibfnamefont {Y.}~\bibnamefont {Cui}}, \bibinfo {author} {\bibfnamefont {Z.}~\bibnamefont {Zhu}}, \ and\ \bibinfo {author} {\bibfnamefont {Q.}~\bibnamefont {Zhan}},\ }\href {\doibase 10.1364/PRJ.6.000138} {\bibfield  {journal} {\bibinfo  {journal} {Photonics Research}\ }\textbf {\bibinfo {volume} {6}},\ \bibinfo {pages} {138} (\bibinfo {year} {2018})}\BibitemShut {NoStop}%
\bibitem [{\citenamefont {Devi}\ and\ \citenamefont {De}(2016)}]{Devi2016}%
  \BibitemOpen
  \bibfield  {author} {\bibinfo {author} {\bibfnamefont {A.}~\bibnamefont {Devi}}\ and\ \bibinfo {author} {\bibfnamefont {A.~K.}\ \bibnamefont {De}},\ }\href {\doibase 10.1364/OE.24.021485} {\bibfield  {journal} {\bibinfo  {journal} {Optics Express}\ }\textbf {\bibinfo {volume} {24}},\ \bibinfo {pages} {21485} (\bibinfo {year} {2016})}\BibitemShut {NoStop}%
\bibitem [{\citenamefont {Devi}\ and\ \citenamefont {De}(2020)}]{Devi2020}%
  \BibitemOpen
  \bibfield  {author} {\bibinfo {author} {\bibfnamefont {A.}~\bibnamefont {Devi}}\ and\ \bibinfo {author} {\bibfnamefont {A.~K.}\ \bibnamefont {De}},\ }\href {\doibase 10.1103/PhysRevResearch.2.043378} {\bibfield  {journal} {\bibinfo  {journal} {Physical Review Research}\ }\textbf {\bibinfo {volume} {2}},\ \bibinfo {pages} {043378} (\bibinfo {year} {2020})}\BibitemShut {NoStop}%
\bibitem [{\citenamefont {Goswami}(2021)}]{Goswami2021}%
  \BibitemOpen
  \bibfield  {author} {\bibinfo {author} {\bibfnamefont {D.}~\bibnamefont {Goswami}},\ }\href {\doibase 10.1088/1742-6596/1919/1/012013} {\bibfield  {journal} {\bibinfo  {journal} {Journal of Physics: Conference Series}\ }\textbf {\bibinfo {volume} {1919}},\ \bibinfo {pages} {012013} (\bibinfo {year} {2021})}\BibitemShut {NoStop}%
\bibitem [{\citenamefont {Bandyopadhyay}\ \emph {et~al.}(2021)\citenamefont {Bandyopadhyay}, \citenamefont {Gaur},\ and\ \citenamefont {Goswami}}]{Bandyopadhyay2021}%
  \BibitemOpen
  \bibfield  {author} {\bibinfo {author} {\bibfnamefont {S.~N.}\ \bibnamefont {Bandyopadhyay}}, \bibinfo {author} {\bibfnamefont {T.}~\bibnamefont {Gaur}}, \ and\ \bibinfo {author} {\bibfnamefont {D.}~\bibnamefont {Goswami}},\ }\href {\doibase 10.1016/j.optlastec.2020.106770} {\bibfield  {journal} {\bibinfo  {journal} {Optics \& Laser Technology}\ }\textbf {\bibinfo {volume} {136}},\ \bibinfo {pages} {106770} (\bibinfo {year} {2021})}\BibitemShut {NoStop}%
\bibitem [{\citenamefont {Zhu}\ \emph {et~al.}(2024)\citenamefont {Zhu}, \citenamefont {Zhang}, \citenamefont {Min}, \citenamefont {Adam}, \citenamefont {Urbach},\ and\ \citenamefont {Yuan}}]{Zhu2024}%
  \BibitemOpen
  \bibfield  {author} {\bibinfo {author} {\bibfnamefont {Z.}~\bibnamefont {Zhu}}, \bibinfo {author} {\bibfnamefont {Y.}~\bibnamefont {Zhang}}, \bibinfo {author} {\bibfnamefont {C.}~\bibnamefont {Min}}, \bibinfo {author} {\bibfnamefont {A.~J.~L.}\ \bibnamefont {Adam}}, \bibinfo {author} {\bibfnamefont {H.~P.}\ \bibnamefont {Urbach}}, \ and\ \bibinfo {author} {\bibfnamefont {X.}~\bibnamefont {Yuan}},\ }\href {\doibase 10.3788/COL202422.023603} {\bibfield  {journal} {\bibinfo  {journal} {Chinese Optics Letters}\ }\textbf {\bibinfo {volume} {22}},\ \bibinfo {pages} {023603} (\bibinfo {year} {2024})}\BibitemShut {NoStop}%
\bibitem [{\citenamefont {Nieto-Vesperinas}\ \emph {et~al.}(2010)\citenamefont {Nieto-Vesperinas}, \citenamefont {Sáenz}, \citenamefont {Gómez-Medina},\ and\ \citenamefont {Chantada}}]{NietoVesperinas2010}%
  \BibitemOpen
  \bibfield  {author} {\bibinfo {author} {\bibfnamefont {M.}~\bibnamefont {Nieto-Vesperinas}}, \bibinfo {author} {\bibfnamefont {J.~J.}\ \bibnamefont {Sáenz}}, \bibinfo {author} {\bibfnamefont {R.}~\bibnamefont {Gómez-Medina}}, \ and\ \bibinfo {author} {\bibfnamefont {L.}~\bibnamefont {Chantada}},\ }\href {\doibase 10.1364/OE.18.011428} {\bibfield  {journal} {\bibinfo  {journal} {Optics Express}\ }\textbf {\bibinfo {volume} {18}},\ \bibinfo {pages} {11428} (\bibinfo {year} {2010})}\BibitemShut {NoStop}%
\bibitem [{\citenamefont {Golat}\ \emph {et~al.}(2023)\citenamefont {Golat}, \citenamefont {Kingsley-Smith}, \citenamefont {Diez}, \citenamefont {Martinez-Romeu}, \citenamefont {Martínez},\ and\ \citenamefont {Rodríguez-Fortuño}}]{Golat2023}%
  \BibitemOpen
  \bibfield  {author} {\bibinfo {author} {\bibfnamefont {S.}~\bibnamefont {Golat}}, \bibinfo {author} {\bibfnamefont {J.~J.}\ \bibnamefont {Kingsley-Smith}}, \bibinfo {author} {\bibfnamefont {I.}~\bibnamefont {Diez}}, \bibinfo {author} {\bibfnamefont {J.}~\bibnamefont {Martinez-Romeu}}, \bibinfo {author} {\bibfnamefont {A.}~\bibnamefont {Martínez}}, \ and\ \bibinfo {author} {\bibfnamefont {F.~J.}\ \bibnamefont {Rodríguez-Fortuño}},\ }\href {http://arxiv.org/abs/2310.11272} {\  (\bibinfo {year} {2023})}\BibitemShut {NoStop}%
\bibitem [{\citenamefont {Toftul}\ \emph {et~al.}(2024)\citenamefont {Toftul}, \citenamefont {Golat}, \citenamefont {Rodríguez-Fortuño}, \citenamefont {Nori}, \citenamefont {Kivshar},\ and\ \citenamefont {Bliokh}}]{Toftul2024}%
  \BibitemOpen
  \bibfield  {author} {\bibinfo {author} {\bibfnamefont {I.}~\bibnamefont {Toftul}}, \bibinfo {author} {\bibfnamefont {S.}~\bibnamefont {Golat}}, \bibinfo {author} {\bibfnamefont {F.~J.}\ \bibnamefont {Rodríguez-Fortuño}}, \bibinfo {author} {\bibfnamefont {F.}~\bibnamefont {Nori}}, \bibinfo {author} {\bibfnamefont {Y.}~\bibnamefont {Kivshar}}, \ and\ \bibinfo {author} {\bibfnamefont {K.~Y.}\ \bibnamefont {Bliokh}},\ }\href@noop {} {\  (\bibinfo {year} {2024})}\BibitemShut {NoStop}%
\bibitem [{\citenamefont {Nelson}\ \emph {et~al.}(2007)\citenamefont {Nelson}, \citenamefont {Li},\ and\ \citenamefont {Weiss}}]{Nelson2007}%
  \BibitemOpen
  \bibfield  {author} {\bibinfo {author} {\bibfnamefont {K.~D.}\ \bibnamefont {Nelson}}, \bibinfo {author} {\bibfnamefont {X.}~\bibnamefont {Li}}, \ and\ \bibinfo {author} {\bibfnamefont {D.~S.}\ \bibnamefont {Weiss}},\ }\href {\doibase 10.1038/nphys645} {\bibfield  {journal} {\bibinfo  {journal} {Nature Physics}\ }\textbf {\bibinfo {volume} {3}},\ \bibinfo {pages} {556} (\bibinfo {year} {2007})}\BibitemShut {NoStop}%
\bibitem [{\citenamefont {Xu}\ \emph {et~al.}(2010)\citenamefont {Xu}, \citenamefont {He}, \citenamefont {Wang},\ and\ \citenamefont {Zhan}}]{Xu2010}%
  \BibitemOpen
  \bibfield  {author} {\bibinfo {author} {\bibfnamefont {P.}~\bibnamefont {Xu}}, \bibinfo {author} {\bibfnamefont {X.}~\bibnamefont {He}}, \bibinfo {author} {\bibfnamefont {J.}~\bibnamefont {Wang}}, \ and\ \bibinfo {author} {\bibfnamefont {M.}~\bibnamefont {Zhan}},\ }\href {\doibase 10.1364/OL.35.002164} {\bibfield  {journal} {\bibinfo  {journal} {Optics Letters}\ }\textbf {\bibinfo {volume} {35}},\ \bibinfo {pages} {2164} (\bibinfo {year} {2010})}\BibitemShut {NoStop}%
\bibitem [{\citenamefont {Vernon}(2024)}]{Vernon2024}%
  \BibitemOpen
  \bibfield  {author} {\bibinfo {author} {\bibfnamefont {A.~J.}\ \bibnamefont {Vernon}},\ }\emph {\bibinfo {title} {Electromagnetic field dark spots}},\ \href@noop {} {Ph.D. thesis},\ \bibinfo  {school} {King's College London} (\bibinfo {year} {2024})\BibitemShut {NoStop}%
\bibitem [{\citenamefont {Griffiths}(2013)}]{Griffiths}%
  \BibitemOpen
  \bibfield  {author} {\bibinfo {author} {\bibfnamefont {D.~J.}\ \bibnamefont {Griffiths}},\ }\href@noop {} {\emph {\bibinfo {title} {Introduction to electrodynamics}}},\ \bibinfo {edition} {4th}\ ed.\ (\bibinfo  {publisher} {Pearson},\ \bibinfo {year} {2013})\ pp.\ \bibinfo {pages} {362--366}\BibitemShut {NoStop}%
\bibitem [{\citenamefont {Bohren}\ and\ \citenamefont {Huffman}(1998)}]{Bohren1998}%
  \BibitemOpen
  \bibfield  {author} {\bibinfo {author} {\bibfnamefont {C.~F.}\ \bibnamefont {Bohren}}\ and\ \bibinfo {author} {\bibfnamefont {D.~R.}\ \bibnamefont {Huffman}},\ }\href {\doibase 10.1002/9783527618156} {\emph {\bibinfo {title} {Absorption and Scattering of Light by Small Particles}}}\ (\bibinfo  {publisher} {Wiley},\ \bibinfo {year} {1998})\BibitemShut {NoStop}%
\end{thebibliography}%

\end{document}